\documentclass{moriond}

\def\be{\begin{equation}}
\def\ee{\end{equation}}
\def\bea{\begin{eqnarray}}
\def\eea{\end{eqnarray}}

\usepackage{amssymb}
\usepackage{amsmath}
\usepackage{mathtools}
\usepackage{floatflt}

\newcommand{\bei}{\begin{itemize}}
\newcommand{\eei}{\end{itemize}}
\newcommand{\bean}{\begin{eqnarray*}}
\newcommand{\eean}{\end{eqnarray*}}

\def\eps{\epsilon}

\def\slh#1{\rlap / {#1}}

\def\slh#1{\rlap / {#1}}

\newcommand{\Photo}{\includegraphics[height=35mm]{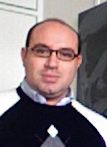}}

\begin{document}
\vspace*{4cm}
\title{QCD Amplitudes: new perspectives on Feynman integral calculus}

\author{Pierpaolo Mastrolia}

\address{
Dipartimento di Fisica e Astronomia, Universit\`a di Padova, and INFN
Sezione di Padova, \\ via Marzolo 8, 35131 Padova, Italy.
\\
Max-Planck Insitut f\"ur Physik, F\"ohringer Ring 6, 80805 M\"unchen, Germany.
}

\maketitle\abstracts{
I analyze the algebraic patterns underlying the structure of
scattering amplitudes in quantum field theory. 
I focus on the decomposition of amplitudes in terms of independent
functions and the systems of differential equations the latter obey.
In particular, I discuss the key role played by unitarity 
for the decomposition in terms of master integrals, 
by means of generalized cuts and integrand reduction, 
as well as for solving the corresponding
differential equations, by means of Magnus exponential series.}

\section{Introduction}
\label{sec:intro}

High energy particle collisions are the ideal framework 
for accessing new informations on matter constituents and forces of nature. 
The higher the energy of the colliding particles, the richer the
landscape of the produced ones. 
The discovery of new physics interactions cannot be disentangled from 
the discovery of massive, {\it heavy} particles, emerging from
collisions of ever increasing energy. On the other side, by increasing
energy, also the probability of producing {\it many light}
particles is enhanced. Therefore, advances in High Energy Particle
Physics necessarily depend on our ability to describe the scattering
processes involving {\it many} light and heavy {\it particles} 
at very high accuracy, hence they depend on our capability of
evaluating Feynman diagrams.
Beyond leading order (LO), Feynman diagrams represent challenging
multidimensional/multivariate integrals, whose direct evaluation is often prohibitive,
therefore the computation of scattering amplitudes beyond the LO is
addressed in two stages: {\it i)} the decomposition in terms of
a basis of functions, and {\it ii)} their evaluation of the elements 
of such a basis, called {\it master integrals} (MIs). 
In this contribution, I elaborate on the algebraic properties of
Feynman integrals, which can be exploited for decomposing them in
terms of MIs and for computing the latter.
The techniques I discuss can be applied to generic amplitudes, and 
have a impact on high-accuracy prediction for collider physics, as well as for
exploring the more formal aspect of quantum field theory.

Let us observe that amplitudes can be decomposed in terms of independent functions, 
exactly like a vector can be decomposed along basic directions. 
One needs a {\it basis} and a {\it projection} technique. 
The latter is necessary to extract the coefficients of the linear
combination.
%
{\it Factorization} is the basic idea we are going to elaborate on. 
Factorization is ubiquitous in the discovery of new mathematical and
physical concepts. 
Complex numbers emerged from factorizing the
simplest number we may think of, {\it i.e.} $1 = (-i) i$;
quantum mechanics relies on the factorization of the identity matrix, 
$\mathbb{I} = \sum_n |n\rangle \langle n|$;
Dirac equation emerged from factorizing the d'Alambertian operator, {\it i.e.} 
$\square = 
(-i \slh{\partial} )
(i \slh{\partial}  )$. 
What does happen when amplitudes factorize?

Cutting a virtual particle and bringing it on the mass shell $(p^2 = m^2)$,
turned out to be a suitable projection technique yielding amplitudes decomposition.
Why multiple-cuts are important? 
First, because multiple-cuts yield functions identification. Since any diagram is characterized by its internal lines, 
a given master diagram is univocally identified by a cut-diagram where all internal particles are on-shell. 
Moreover, when applied to amplitudes, multiple-cuts behave like {\it
  high-pass filters}, 
which isolate only the diagrams that have those
internal lines to be cut, while the others are automatically discarded. 
Therefore, by considering all possible cuts of an amplitude, in a
top-down procedure, from the maximum number of cuts \cite{} to the lowest one,
it is possible to build a (triangular) system of equations from which
{\it all} coefficients can be determined. 

\section{Integrand decomposition}

\begin{floatingfigure}[l]{5.0cm}
\hspace*{-1cm}
\includegraphics[width=0.35\textwidth]{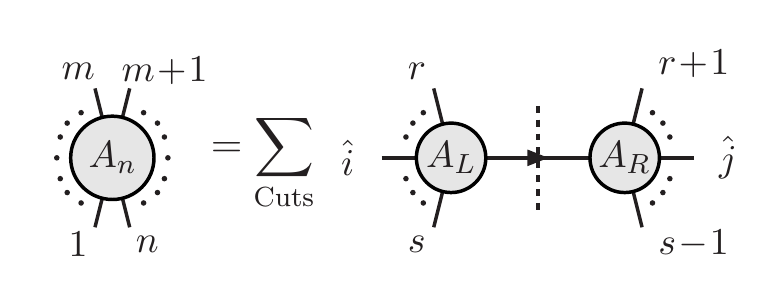}
\caption{Tree-level recurrence relation.}
\label{Fig:BCFW}
\end{floatingfigure}
Tree-level scattering amplitudes obey a quadratic
recurrence relation \cite{Britto:2004ap} (BCFW), depicted in Fig.\ref{Fig:BCFW}, whose derivation relies on  Cauchy's
residue theorem.
Since tree amplitudes are rational functions of kinematic variables,
the BCFW recurrence can be understood as due 
simply to partial fractioning~\cite{Vaman:2005dt}, 
because {\it residue theorem applied to rational functions amounts to
partial fractions}.
Is that just accidental, and holding for tree-level amplitudes, or partial fractioning can be exploited also
at higher orders?

The {\it integrand reduction algorithm} \cite{Ossola:2006us} 
had a dramatic impact on our ability of computing %
\begin{floatingfigure}[r]{7.0cm}
\hspace*{-1cm}
\includegraphics[width=0.45\textwidth]{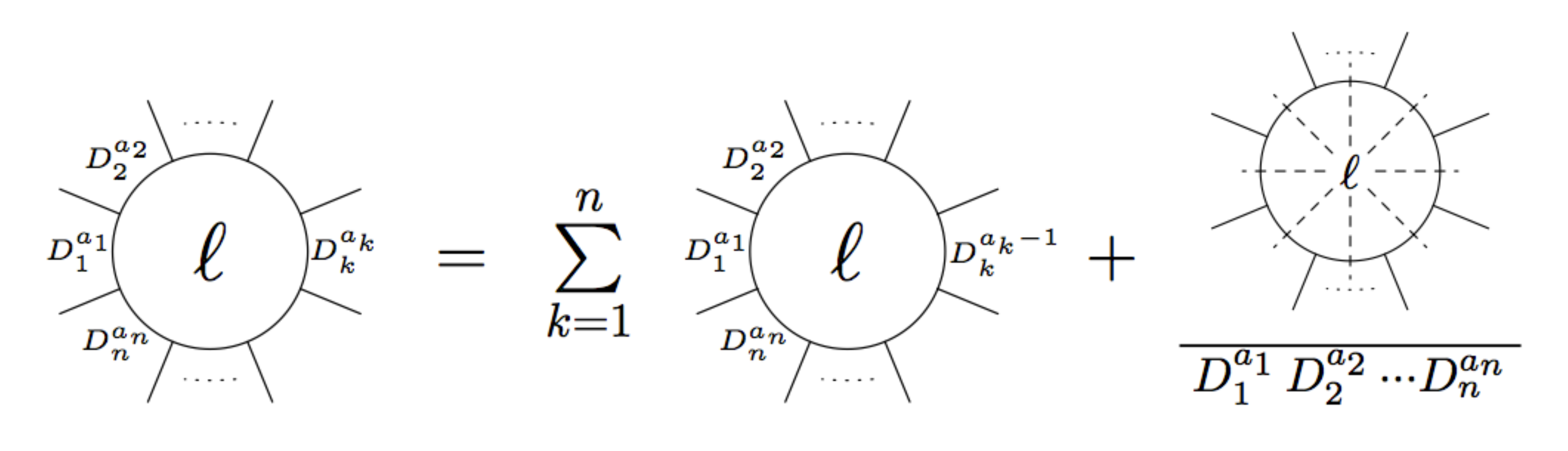}
\caption{Multiloop integrand decomposition. 
}
\label{Fig:IntRed}
\end{floatingfigure}
\noindent
one-loop amplitudes.
The basic idea lies in the existence of a relation between numerators and denominators
of scattering amplitudes which can be used to decompose the integrands
of one-loop amplitudes in terms of integrands of MIs. The amplitude
decomposition in terms of MIs is then 
achieved after integrating the integrand decomposition. 
The coefficients of the MIs are a subset of the coefficients
appearing in the decomposition of the integrands. Therefore, within 
the integrand reduction algorithm, coefficients can be determined
simply by algebraic manipulation, 
with the great advantage of bypassing any integration. 

\begin{floatingfigure}[l]{5.5cm}
\hspace*{-8cm}
\includegraphics[width=0.33\textwidth]{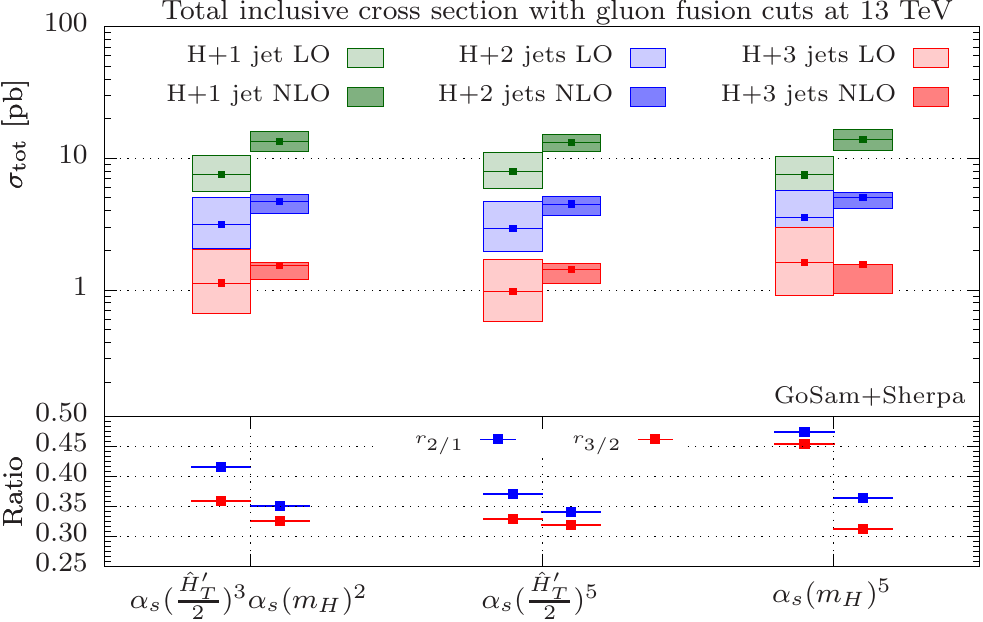}
\caption{\footnotesize LO and NLO total cross sections for $pp \to H+$1,2,3,jets 
at LHC (\!13 TeV\!).}
\label{Fig:H123j}
\end{floatingfigure}
\noindent
The idea behind the {\sc GoSam} framework \cite{Cullen:2011ac,Cullen:2014yla} 
is to combine automated diagram
generation and algebraic manipulation with the integrand-level
reduction, implemented 
in {\sc Samurai}~\cite{Mastrolia:2010nb,vanDeurzen:2013pja} and 
{\sc Ninja}~\cite{Mastrolia:2012bu,Peraro:2014cba}
and the tensor decomposition of {\sc Golem95}~\cite{Guillet:2013msa}.
The code is very flexible and it has been employed in several applications at NLO QCD
accuracy, studies of BSM scenarios, electroweak calculations, and
recently also within NNLO calculations. It is interfaced to several MonteCarlo
event generators, like {\sc Sherpa}, {\sc Herwig}, {\sc aMC@NLO}.
{\sc GoSam} was used to evaluate the NLO QCD correction to $pp \to
Hjj, Hjjj$ 
(in the infinite top-mass limit) \cite{vanDeurzen:2013rv,Cullen:2013saa}, which required an extension of the
integrand decomposition methods \cite{Mastrolia:2012bu}.
The evaluation of the virtual amplitudes for $pp \to Hjjj$  has been further optimized,
enhancing the numerical accuracy and reducing the 
computing time \cite{vanDeurzen:2013saa}, ending up into a new phenomenological analysis~\cite{Deurzen:2014mda,Greiner:2015jha}, 
obtained with the tandem of {\sc GoSam} and {\sc Sherpa}, see Fig.~\ref{Fig:H123j}.

The extension of the integrand decomposition beyond one-loop has been proposed in \cite{Mastrolia:2011pr}, 
and refined in \cite{Badger:2012dp,Zhang:2012ce,Mastrolia:2012an},
where the unitarity-based decomposition of multi-loop integrands has
been addressed as a polynomial decomposition problem, and systematized
within the {\it multivariate polynomial division algorithm}.
Accordingly, any generic multi-loop integral with $n$ denominators,
$
{\cal I}_{12\ldots n} = \int d^d q_1 \cdots d^d q_m \ I_{12\ldots n} ,
$ with
$
I_{12\ldots n} = {N_{12\ldots n} /(D_1 \ldots D_n)} ,
$
can undergo an integrand decomposition 
by means of successive polynomial divisions (modulo Gr\"obner basis)
between the numerator and the denominators,
see Fig.\ref{Fig:IntRed}.
The result of the decomposition reads as,
\bea
I_{12\ldots n} &=& 
{\Delta_{12\ldots n} \over D_1 \ldots D_n} 
+ {\Delta_{2\ldots n} \over D_2 \ldots D_n} 
+ \ldots 
+ {\Delta_{12\ldots n-1} \over D_1 \ldots D_{n-1}} 
+ \ldots 
+ {\Delta_{ n} \over D_n} 
+ \ldots 
+ {\Delta_{ 1} \over D_1} \ ,
\label{eq:intdeco}
\eea
where $\Delta_{i \ldots j}$ are the remainders of the iterated divisions
(w.r.t. the  Gr\"obner basis of the ideal $\langle D_i, \ldots, D_j \rangle$).
Each residue $\Delta_{i \ldots j}$ is a polynomial in the components
of the loop momenta not constrained by the cut $ D_i = \ldots = D_j =0$.
Therefore, by integrating both sides, one obtains the decomposition of the original
integral ${\cal I}_{12\ldots n}$ in terms of {\it independent integrals}.
The integrand decomposition (\ref{eq:intdeco}) implies that, exactly
as it happens for the tree-level amplitudes, also the integrands of
multi-loop amplitudes can be decomposed in terms of independent
building blocks simply by {\it partial fractioning} !

While in the one-loop case the independent integrals are analytically known, 
in the multi-loop case, their classification and evaluation is an open
problem. \\ \\

\section{Differential Equations and Feynman Integrals}
\label{sec:diffeq}

The {\it method of differential equations} (DEs)
\cite{Kotikov:1990kg,Remiddi:1997ny,Gehrmann:1999as}, 
reviewed in \cite{Argeri:2007up,Smirnov:2012gma,Henn:2014qga}, is one of the most effective
techniques for computing dimensionally regulated multi-loop integrals.
\begin{floatingfigure}[r]{5.0cm}
\includegraphics[width=0.20\textwidth]{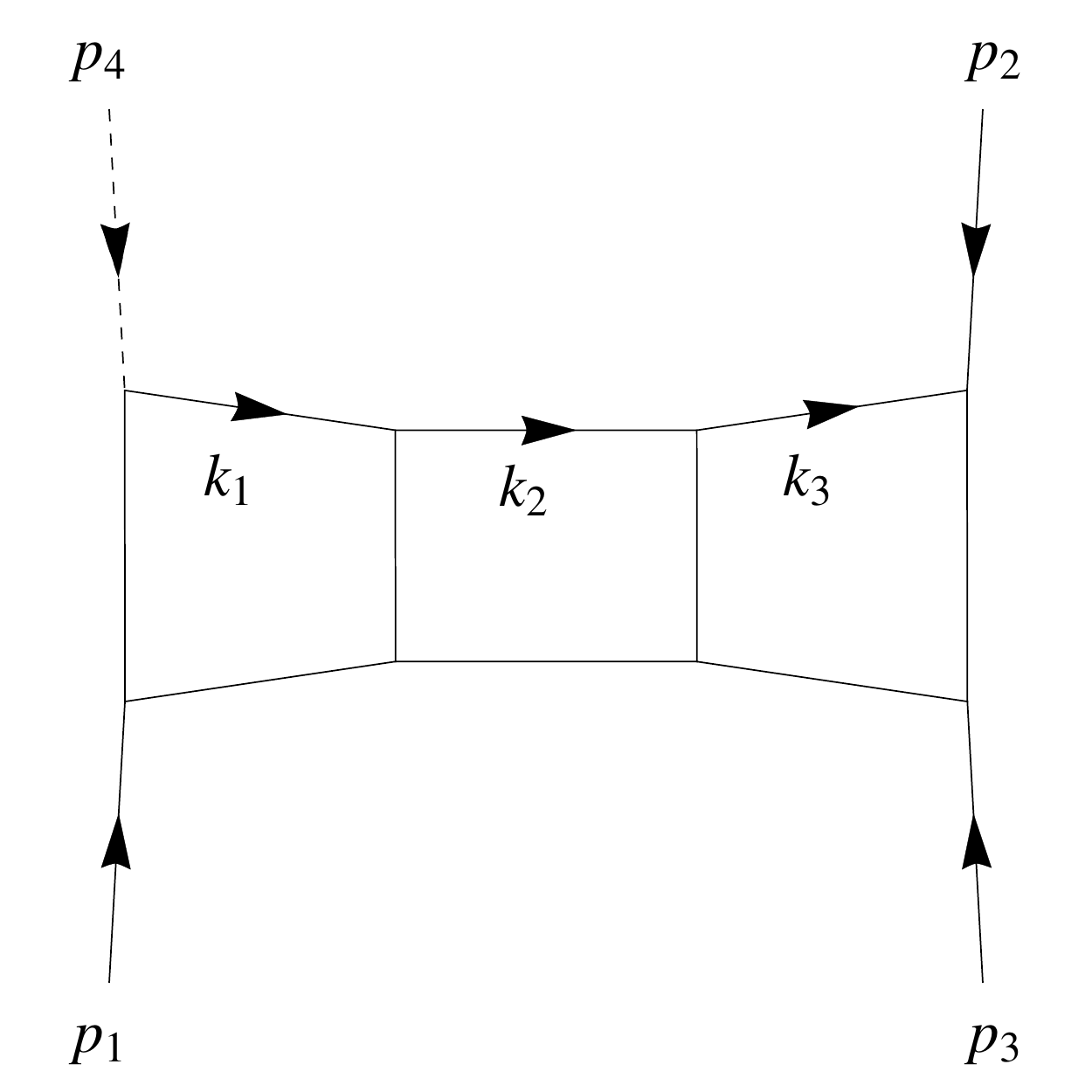}
\caption{The three-loop
ladder box diagram, with one off-shell leg (dashed line)
}
\label{Fig:3L1mLadder}
\end{floatingfigure}
In fact, any $\ell$-loop integral ${\cal I}$ is a homogeneous function
of external momenta $p_i$ and masses $m_i$, whose degree $\gamma =
\gamma(d,\ell)$ 
depends on the space-time dimensions $d = 4-2\eps$, on the number of loops $\ell$,
and on the powers of denominators. Therefore, one can write the
Euler scaling equation,
\bea
\Big(\sum_i p_i \cdot \partial_{p_i} + 
\sum_j m_j^2  \ \partial_{m_j^2}\Big) {\cal I} = \gamma(d,\ell) {\cal I} \ ,
\eea
where $\partial_x \equiv \partial/\partial x$.
Euler relation can be engineered to show that MIs obey linear systems of first-order differential equations (DEs) in
the kinematic invariants,
which can be used for the determination of their actual 
expression. 
By establishing an analogy between Schr\"odinger Equation in the
interaction picture 
(in presence of an Hamiltonian with a linear perturbation) and 
systems of DEs for Feynman integrals (whose associated matrix is
linear in $\eps$)~\cite{Argeri:2014qva},
we have recently proposed an algorithm to find the transformation matrix
yielding to a {\it canonical} system~\cite{Henn:2013pwa}, where the dependence on the dimensional parameter 
$\epsilon$ is factorized from the kinematic. 
In particular, we found that the canonical transformation can be
obtained by means of Magnus exponential matrix \cite{Magnus}.
The integration of canonical systems is simple, 
and the analytic properties of its solution are manifestly
inherited from the associated matrix, that becomes the kernel of the
representation of the solutions in terms of repeated integrations. 
The latter in fact are the coefficients of a
Magnus (or alternatively Dyson) series expansion in $\epsilon$.
Magnus exponential is not unitary, as it happens in the quantum
mechanical case, 
but the proposed method can be considered also inspired by unitarity.

\subsection{Applications}

We made use of Magnus theorem for the determination of non-trivial
integrals, like the two-loop 
\begin{floatingfigure}[r]{5.0cm}
\hspace*{-0.5cm}
\includegraphics[width=0.30\textwidth]{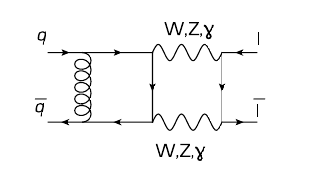}
\vspace*{-.5cm}
\caption{\footnotesize Representative two-loop
box diagram for QCD-EW corrections to $q  {\bar q} \to \ell^+ \ell^-$.}
\label{Fig:DY}
\end{floatingfigure}
\noindent
vertex diagrams for the electron form
factors in QED and the two-loop box integrals for the $2 \to 2$
massless scattering \cite{Argeri:2014qva}, 
the two-loop corrections to the $pp \to Hj$, as well as for evaluating 
the three-loop ladder diagrams for $pp \to Hj$ (in the infinite
top-mass approximation) \cite{DiVita:2014pza}, see
Fig.\ref{Fig:3L1mLadder}. 
The latter is a formidable calculation involving the solution of a system of
85 MIs. 
In this case, after identifying a set of MIs obeying
a linear system of differential equations in $x= - s/m_H^2$ and $y=
-t/m_H^2$, by means of a Magnus transform, 
the system can be brought in canonical form, reading as,
\bea
d {\vec{\cal I}} (x,y) = \eps \ A(x,y) \ {\vec{\cal I}} (x,y) \ , 
\eea
where ${\vec{\cal I}} $ is the vector of MIs, 
and $df \!=\! \partial_x f dx + \partial_y f dy$. The matrix $A$ is purely logarithmic, 
$
A(x,y) =
  a_1 \ln(x)
+ a_2 \ln(1-x) 
+ a_3 \ln(y) 
+ a_4 \ln(1-y) 
+ a_5 \ln(x+y)
+ a_6 \ln(1-(x+y)) 
\ , $
where the $a_i \ (i=1,\ldots,6)$ are $85 \times 85$ matrices whose
entries are just rational numbers.
The logarithmic form of $A$ trivializes the solution, which can be 
written as a Dyson series in $\eps$, where the coefficient of the
series are combinations of Multiple Polylogarithms with uniform weight (where the weight increases as the
order in $\eps$ does). 
\begin{floatingfigure}[r]{8.0cm}
\hspace*{-0.5cm}
\includegraphics[width=0.5\textwidth]{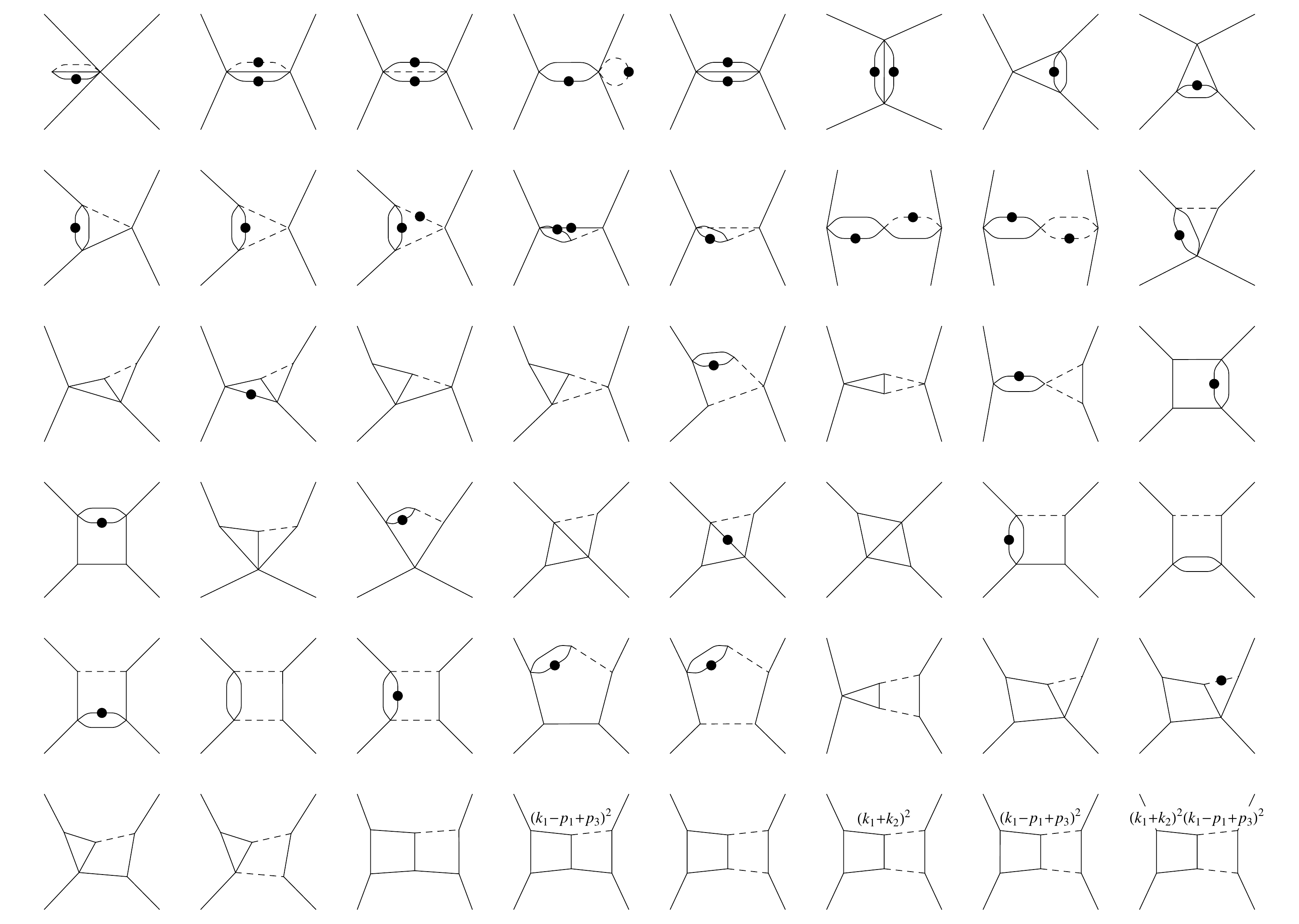}
\caption{\footnotesize Master integrals for $q  {\bar q} \to \ell^+
  \ell^-$ at two-loop. Plane lines stands for massless particle, while
dashed lines stands for massive particles. }
\label{Fig:DY:MIs}
\end{floatingfigure}
Boundary conditions are determined by imposing the regularity of the
solutions in special kinematic configurations. Surprisingly, to fix the
boundary values of all 85 MIs, only 2 simple integrals have to be
independently provided. \\ 

Also, we have been considering the mixed EW-QCD corrections to Drell-Yan
production at NNLO, whose representative diagram is depicted in
Fig.{\ref{Fig:DY}}. Also in this case, Magnus
exponential can employed to reach a canonical system 
for the 48 MIs drawn in Fig.\ref{Fig:DY:MIs}, in the variables $x= -
s/m_V^2$ and $y=-t/m_V^2$, with $V=W,Z$ \cite{BDMS}.

\section{Conclusions}

In this contribution, I have analyzed the algebraic patterns underlying
the structure of scattering amplitudes in gauge theory. 
{\it Unitarity} plays a central role in
the context of evaluating scattering amplitudes.
It not only inspired methods to perform the amplitudes
decomposition, by means of unitarity-cuts, but it also suggested a
technique for the evaluation of master integrals, by means of
matrix exponentials, similar to the unitary time-evolution in quantum
mechanics. 

\section*{Acknowledgments}
Supported by the Alexander von Humboldt Foundation,
in the framework of the Sofja Kovalevskaja Award 2010, endowed by the
German Federal Ministry of Education and Research.

\section*{References}

\bibliography{references}

\end{document}